\documentstyle[psfig,rotate]{article}
\def\bea{\begin{eqnarray}}
\def\eea{\end{eqnarray}}
\def\bq{\begin{quote}}
\def\eq{\end{quote}}

\def\gappeq{\mathrel{\rlap {\raise.5ex\hbox{$>$}}
{\lower.5ex\hbox{$\sim$}}}}

\def\lappeq{\mathrel{\rlap{\raise.5ex\hbox{$<$}}
{\lower.5ex\hbox{$\sim$}}}}

\def\Toprel#1\over#2{\mathrel{\mathop{#2}\limits^{#1}}}

\begin{document}
\pagestyle{empty}
\begin{flushright}
{CERN-TH/2000-253}
\end{flushright}
\vspace*{5mm}
\begin{center}
{\Large \bf STRANGE HADRONS FROM QUARK AND HADRON MATTER: A COMPARISON
} \\
\vspace*{1cm} 
{\bf T.S. Bir\'o$^{1,2}$, P. L\'evai$^2$ and J. Zim\'anyi$^2$} \\
\vspace{0.3cm}
$^1$ Theoretical Physics Division, CERN, CH - 1211 Geneva 23 \\
$^2$ MTA KFKI, H-1525 Budapest P.O.B. 49, Hungary \\

\vspace*{15mm}  
{\bf ABSTRACT} \\ \end{center}
\vspace*{5mm}
\noindent

In a simplified model we study the hadronization of
quark matter in an expanding fireball, 
in particular the approach to  a final hadronic composition
in equilibrium.  
Ideal hadron gas equilibrium
constrained by conservation laws, the fugacity parametrization,
as well as linear and non-linear coalescence approaches
are recognized as different approximations to this 
in-medium quark fusion process.
It is shown that color confinement requires a dependence of the hadronization
cross section on quark density in the presence of flow (dynamical confinement).

\vspace*{40mm} 

\centerline{{\em To appear in the Proc. of the Strange Matter 2000 Conference}}

\noindent 

\noindent
\vspace*{40mm}

\begin{flushleft} CERN-TH/2000-253 \\
August 2000
\end{flushleft}
\vfill\eject

\setcounter{page}{1}
\pagestyle{plain}



%
%
%


\newcommand{\insertplot}[1]{
\begin{center}\leavevmode\epsfysize=13.0cm \epsfbox{#1}\end{center}}

\newcommand{\be}{\begin{equation}}
\newcommand{\ee}{\end{equation}}
\newcommand{\vs}{\vspace{0mm}}




\vs
\section{Introduction}

\vs
Since flavor composition, in particular strangeness enhancement 
had been suggested as a possible signature
of quark matter formation in heavy-ion collisions, there is
a continuing attention to the question of hadronic and quark level
equilibrium in these reactions 
\cite{Rafelski:1982pu,Biro:1982zi}. 
Recent claims of describing data
at AGS as well as at CERN SPS by hadronic equilibrium
(i.e. using a common volume, temperature and only chemical
potentials of conserved quantities for the computation of hadron
numbers) satisfactorily 
\cite{Braun-Munzinger:1999qy,Becattini:1998we,Cleymans:1999yf} 
led to some discrepancy:
the total reaction time at SPS energy is 
too short for achieving hadronic equilibrium by hadronic conversion
processes alone.
Experimental data are well described, but poorly understood.
After the ``go and find it'' period now we are at the
``sit down and think'' stage.

\vs
There are also several non-equilibrium approaches
to this problem ranging from the use of phenomenological
parameters on the hadronic level (fugacities) 
\cite{Letessier:1999sz}
through linear \cite{Bialas:1998ea} and nonlinear 
(constrained) coalescence models
\cite{Biro:1995mp},
and mixed quark-hadron chemistry \cite{Biro:1999cx} 
up to very detailed microscopic simulations of the hadronization
(parton cascades \cite{Geiger:1993cm,Geiger:1997ym}
VENUS \cite{Werner:1987ze},
HIJING \cite{Eskola:1996bp},
RQMD \cite{Sorge:1995vv,Sollfrank:1999zg},
UrQMD \cite{Antinori:1999rw,Bleicher:1999cw}, 
string model
\cite{Amelin:1990vp,Dean:1992vj,Bass:1998ca},
chromodielectric model \cite{Traxler:1999bk}).
The basic physical concepts of these approaches do not include  hadronic
equilibrium; it might only be reached in course of time. 

\vs
The purpose of this talk is to 
understand the near-equilibrium hadronic result in terms of
quark number evolution. We consider the simplest possible
process on the constituent level: 
$ A + B \longleftrightarrow C.$

\vs
In such a process the energy-impulse balance is assumed to be satisfied by the
surrounding medium, we are interested in the time evolution of
the number of final state products $N_C(t)$. The equilibrium number
of precursors $N_{A,eq}$ and $N_{B,eq}$ - in a quark matter scenario - will 
be taken as zero. The initial state is assumed to be far from this
equilibrium, it is oversaturated by $A$ and $B$ with
no $C$ particles at all in the beginning.
As the reaction proceeds there will be $C$ type particles
(hadrons) produced.
The cross section for such a reaction at a first glance
can be taken as  constant, but temperature and reaction volume
must be considered as time dependent quantities.
We shall point out, however, that the assumption of constant
cross sections actually contradicts to the confinement principle
in quark matter scenarios with transverse flow, so in that case the hadronization
cross section has to be quark density dependent. This extra dependency,
not derivable from perturbative or lattice QCD so far, will be referred
as ``dynamical confinement'' in this paper.


\vs
\section{Chemistry}

\vs
In the chemical approach to hadronization 
all particle numbers are time dependent.
A general reaction can be characterized by stochiometric 
coefficients $\alpha_i^{(r)}$ 
describing particle number changes in a given reaction $(r)$
in a fashion arranged to zero:
\be
 \sum_i \alpha_i^{(r)} A_i = 0,
\label{STOCHIO}
\ee
where $A_i$ symolically denotes one particle of type $i$.
In this arrangement the $\alpha_i^{(r)}$-s 
can be negative as well as positive integers.
(For the simple quark fusion process $A+B\longleftrightarrow C$
one applies $\alpha_A = \alpha_B = 1, \alpha_C=-1$.)

\vs
The decisive quantity with respect to approaching chemical equilibrium
is the activity of a reaction,
\be x^{(r)} = \sum_i \alpha_i^{(r)} \frac{\mu_i}{T}.\ee 
It describes the (un)balance of gain and loss terms.
Rate equations add up  contributions from several reaction channels,
\be 
 \dot{N}_i = \sum_{(r)} \alpha_i^{(r)} R^{(r)} 
 \left( 1 - e^{x^{(r)}} \right). 
 \label{RATE_EQ}
\ee
Here the factors $R^{(r)}$ depend on numbers of other particles,
cross sections and volume. Due to known (e.g. adiabatic) expansion
a relation between volume $V$ and temperature $T$ is given.
In chemical equilibrium $ x^{(r)} = 0 $ for each reaction, 
\be \sum_i \alpha_i^{(r)} \mu_i = 0, \ee
so there are linear constraints for the chemical potentials $\mu_i$ 
as many as reaction channels.
If this is less than the number of particle sorts,
the equilibrium state is undetermined.
If this is greater than the number of particle sorts,
the reactions cannot be all independent: there are conserved
quantities. 
In this latter case all $\mu_i$ can be expressed by a few numbers
corresponding to these conserved quantities. This is the constrained
equilibrium.

\vs
Denoting the conserved quantities by $Q_a$ and
the corresponding charge of the i-th particle sort by $q_{i,a}$,
we obtain
\be \sum_i q_{i,a} \dot{N}_i = 0. \ee
Substituting the rate equations (\ref{RATE_EQ}) we get
\be \sum_i \sum_r \alpha_i^{(r)}q_{i,a} R^{(r)} \left(1-e^{x^{(r)}} \right)
= 0. \ee
Since this holds at any instant of an out of equilibrium evolution,
the double sum vanishes for arbitrary values of reaction activities
$x^{(r)}$. Exchanging the order of summation over particle types $i$
and reaction channels $r$, and taking into account that all $R^{(r)}$
factors are positive, we conclude that
the considered charges are conserved in each reaction:
\be 
  \sum_i \alpha_i^{(r) } q_{i,a} = Q^{(r)}_a = 0. 
  \label{CONSERV}
\ee
In this case we can redefine the chemical potentials as
\be \mu_i = \sum_a q_{i,a} \mu_a + \tilde{\mu}_i,\label{REDEF-MU} \ee 
without changing the activities:
\be x^{(r)} = \sum_i \sum_a \alpha_i^{(r)} \frac{q_{i,a}\mu_a}{T}
+ \sum_i \alpha_i^{(r)} \frac{\tilde{\mu}_i}{T}. \ee
Since the first sum vanishes due to eq.(\ref{CONSERV}), 
chemical equilibrium is equivalent to
\be \sum_i \alpha_i^{(r)} \frac{\tilde{\mu}_i}{T} = 0. \ee
For a well-determined or overdetermined system the only equilibrium
solution is given by $\tilde{\mu}_i=0.$
In this (constrained) equilibrium state the original chemical potentials
are a linear combination of a few $\mu_a$-s, associated to conserved charges,
(cf. eq.(\ref{REDEF-MU}))
\be \mu_{i,eq} = \sum_a q_{i,a} \mu_a. \ee 
Whether this chemical equilibrium state can be reached in a finite time
depends on
the rates $R^{(r)}$ and activities $x^{(r)}$.


\vs
\section{A simple model: A + B $\longleftrightarrow$ C }

\vs
In the case of only one reaction channel,
$ A + B \longleftrightarrow C,$ we denote the equilibrium ratio by
\be
R_{{\rm eq}} = \frac{N_{A,{\rm eq}}N_{B,{\rm eq}} }{ N_{C,{\rm eq}} }.
\label{EQUIL-RATIO}
\ee
Since $\alpha_A=\alpha_B=1$ and $\alpha_C=-1$ for this very reaction,
we have $\dot{N}_A=\dot{N}_B=-\dot{N}_C$. From this two
conservation laws follow,
\be N_A + N_C = N_A(0) = N_0, \label{CONA} \ee
\be N_B + N_C = N_B(0) = N_0. \label{CONB} \ee
For considering an evolution towards equilibrium the
specific rate,
$ \nu =  \langle \sigma v \rangle/V,$
is a useful quantity. It occurs in the
rate equation 
\be 
   \dot{N}_C = \nu \left( N_A N_B - R_{eq} N_C  \right). 
\ee
Substituting the conservation laws (\ref{CONA},\ref{CONB}) we arrive at
\be 
  \dot{N}_C = \nu \left((N_0-N_C)(N_0-N_C)- R_{eq} N_C \right).
  \label{SIMPLE_RATE_EQ}
\ee
This rate is vanishing for the particle numbers $N_C=N_{\pm}$ with
\be N_{\pm} = R_{eq}/2 + N_0 \pm \sqrt{R_{eq}^2/4 + R_{eq}N_0}. \ee
Here $N_-$ is a stable, $N_+$ is an unstable equilibrium point.
For an adiabatic expansion
$R_{eq} = const.$ for constituent hadrons 
(those with mass $m_C = m_A + m_B$).
The simple rate equation (\ref{SIMPLE_RATE_EQ}) can be solved
analytically with the result 
\be  
   \int_{N_C(0)}^{N_C(\infty)} \frac{dN_C}{(N_C-N_+)(N_C-N_-)} =
   \int_{t_0}^{t_{br}} \nu dt = r. 
\ee
Here $t_0$ is the initial time, when the fireball contains no hadrons,
$N_C(0)=0$. $t_{br}$ is the break-up time, beyond which no chemical
reactions (even no particle collisions) occur. In some cases it is
large compared to characteristic times and can be approximated by
infinity.
The solution can be written as
\be 
   N_C(\infty)= N_+N_- \frac{1-K}{{N_+} - K {N_-}} 
\ee
with
\be K =  \exp \left( -  (N_+-N_-)r \right). \ee
Here $N_-$ is the constrained equilibrium value, achieved for $K=0$.
There are two alternative forms of this relation 
derived by using $N_+N_-=N_0^2$.
The first one is the coalescence form,
\be N_C(\infty) = N_0^2 \frac{1-K}{N_+-KN_-} = 
C(K,N_0) N_A(0) N_B(0),  \ee
the second one is the equilibrium form,
\be N_C(\infty) = N_- \frac{1-K}{1-K(N_-/N_0)^2} 
= \gamma_C N_{C,{\rm eq}}. \ee
In particular $ N_C(\infty) = N_-$ in the $K=0$ (infinite rate) case.
On the other hand in the case of confinement $R_{eq}=0$, from which
$N_+=N_-=N_0$  and $K=1$ follow. 

\vs
For $(N_+-N_-)$ 
small (${\cal O}(R_{eq})$), but non-zero the precursors 
are already hadronic, but suppressed in the equilibrium state. 
Then we expect, that in the $R_{{\rm eq}} \rightarrow 0$ limit 
the precursors
$A$ and $B$ are no more present in the final mixture.
This is, however, insufficient
for the quark matter hadronization scenario,
because the above limit does not automatically lead
to an absence of $A$ and $B$ particles in the final state,
if the expansion is dramatic enough.
We describe this phenomenon in the followings.

\vs
Let us expand $K$ in terms of $(N_+-N_-)$ which is in the same order of
magnitude as $R_{{\rm eq}} \rightarrow 0$. We get
\be K = \exp(- (N_+-N_-)r) \approx 1 - r(N_+-N_-), \ee
and the final hadron number becomes
\be 
  N_C(\infty) = N_0 \frac{rN_0}{1+rN_0},
  \label{BASIC}
\ee
leaving us with a non-vanishing precursor number
\be
  N_A(\infty) = N_B(\infty) = N_0 \frac{1}{1+rN_0}.
\ee

\vs
Eq. (\ref{BASIC}) has different readings. 
For small $r$  it is close to the result of linear 
coalescence models
\be
  N_C(\infty) = r N_0^2.
  \label{LINEAR}
\ee
(Note that $N_A(0)=N_B(0)=N_0$.)
ALCOR uses extra modification factors $b_A$ and
$b_B$ in the coalescence rate
\be
  N_C(\infty) = r b_A b_B N_0^2.
  \label{NON_LINEAR}
\ee
To start with it seems to be rather similar to the coalescence picture.
Due to conservation of quark number during an -- supposedly rapid --
hadronization, however, ALCOR obtains
\be
  N_C(\infty) = r b_A b_B N_0^2 = \frac{r}{1+rN_0}N_0^2
  \label{ALCOR}
\ee
and determines the factors $b_i$ accordingly 
\be
  b_A = b_B = \frac{1}{\sqrt{1+rN_0}}.
\ee
It leads exactly to the constrained equilibrium result in the
infinite integrated specific rate limit $rN_0 \rightarrow \infty$
(then $b_A=b_B \approx 1/\sqrt{rN_0}$):
\be
  N_C(\infty) =  N_0, \qquad \qquad 
  N_A(\infty) =  N_B(\infty) = 0.
  \label{EQUIL}
\ee
In this limit the precursor numbers approach zero, and the above
discussion can be applied for quark matter hadronization.
Or, reversing the argument, we conclude that {\em for quark matter hadronization
$r$ has to diverge} in order to agree with confinement in the end stage.
(This fact is nonetheless hidden in the thermodynamical limit with infinite
quark number $N_0 \rightarrow \infty$.)

\vs
Finally the chemistry approach gives the interpretation of
the fugacity due to
\be
  N_C(\infty) = N_0^2 \frac{r}{1+rN_0} = \gamma_C N_0.
  \label{FUGA}
\ee
In fact the relationship between these approaches is more pronounced
by inspecting chemical potentials instead of final numbers.
In the discussed simple case the chemical potential of the
(final state) hadron is given by
\be
 \frac{\mu_C}{T} = \log \frac{N_C(\infty)}{N_{C,eq}}.
 \label{BOLTZMANN_CHEM_POT}
\ee
Substituting our result for $N_C(\infty)$ it splits to contributions
with different physical meanings:
\be
 \frac{\mu_C}{T} = \log \frac{N_-}{N_{C,eq}} \, + \, 
  \log \frac{N_C(\infty)}{N_-}
 = \frac{q_{C,1}\mu_1 + q_{C,2}\mu_2}{T} + \log \gamma_C.
 \label{CHEM_POT_C}
\ee
The conserved quantities (charges) carried by the hadron $C$ 
are those involved in the identities respected by the rate equations,
$N_A(t)+N_C(t)=N_A(0)=N_0$ and $N_B(t)+N_C(t)=N_B(0)=N_0$. In this case 
$q_{C,1}=1$ and $q_{C,2}=1$, respectively. The fugacity parameter
(as used by Rafelski et al.) is the remaining part of the formula,
in equilibrium it approaches one: $\gamma_C=1$.


\vs
We note here that the ALCOR approach (constrained coalescence)
leads to the equilibrium result only if a single reaction channel
is considered, like in the case discussed in the present paper.
For several, competing channels the relative hadron ratios
are different for ALCOR and constrained equilibrium.
This best can be seen by inspecting the corresponding $\mu_i$
chemical potentials to the pre resonance decay hadron numbers
as a function of strangeness in the ALCOR simulation.
Slight deviations from the
perfect straight lines, which would mark the constrained equilibrium
in the ideal case, can be observed 
in particular for multiply strange anti-baryons
(cf. Fig.1).

\begin{figure}

\centerline{\psfig{file=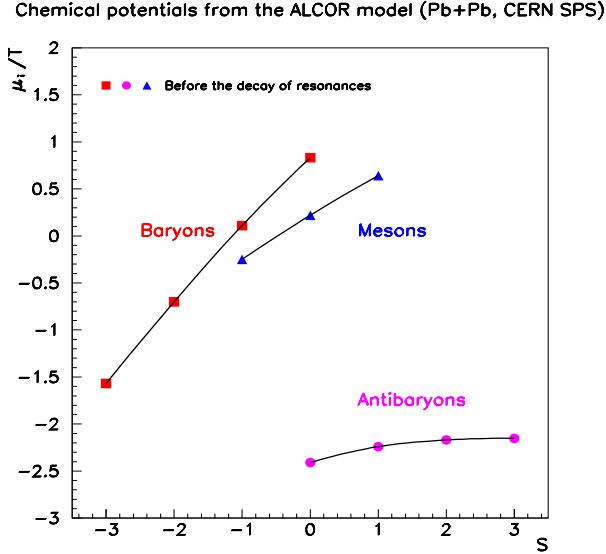,width=75mm,height=100mm}}
\label{ALCOR-MU}

\vspace*{-10mm}
\caption{
Equivalent chemical potentials of composite hadrons before  
resonance decay 
obtained in ALCOR as a function of the strangeness. Linearity corresponds
to hadronic equilibrium.
}

\end{figure}


\section{Expansion effects}

\vs
We continue by obtaining the integrated specific rate, $r,$ for
different expansion scenarios.
This quantity depends not only on cross section(s) -- first assumed
to be constant and later density dependent -- but also on the
expansion and cooling of the reaction zone. We consider here two
cases: one dimensional expansion with a
rather relativistic equation of state  
and  three dimensional
expansion of a massive, non-relativistic ideal Boltzmann gas. The first
case is intended to be characteristic for RHIC and LHC, the
latter rather for $SPS$ and $AGS$ energies.

\vs
During a one-dimensional expansion the volume grows linearly in time,
\hbox{$ V = \pi R_0^2 t$,}
and the (thermal) average of relative velocities is near to the
light speed $\langle v \rangle = 1$. In this case we obtain
\be
 r = \int_{t_0}^{t_{br}} \frac{\sigma}{\pi R_0^2} \frac{dt}{t}
 = \frac{\langle \sigma v_0 \rangle}{V_0} t_0\log \frac{t_{br}}{t_0}.
 \label{RHIC_rate}
\ee
The time-integrated specific rate grows with the break-up time $t_{br}$
monotonically, it can eventually be infinite, if there is no thermal break-up
of the expanding system. In that case $\gamma_C=1$, equilibrium
is achieved.
For a three-dimensional expansion one gets $V=V_0(t/t_0)^3$. Furthermore
a massive, non-relativistic ideal gas satisfies $VT^{3/2}=V_0T_0^{3/2}$
during an adiabatic expansion. The average relative velocity is 
temperature dependent like 
\hbox{$\langle v \rangle \sim \sqrt{T/m} \sim 1/t$,}
with some reduced mass of the fusing pair $m$. This together leads to
the estimate
\be
  r = \frac{\langle \sigma v_0 \rangle}{V_0} \int_{t_0}^{t_{br}}
t_0^4 \frac{dt}{t^4}.
\ee
Evaluating the integral we arrive at
\be
 r = \frac{1}{3} \frac{\langle \sigma v_0 \rangle}{V_0} t_0 \left( 
 1 - \left( \frac{t_0}{t_{br}}\right)^3 \right).
 \label{SPS_rate}
\ee
It is important to note that {\em $r$ is finite even without break-up,}
in the $t_{br} \rightarrow \infty$ limit.
This means that the non-relativistic, 3-dimensional spherical
expansion scenario with constant hadronization cross section cannot
comply with quark confinement. 
This result is valid in the presence of any, however mild, transverse
flow, since the one-dimensional expansion scenario showed a
logarithmic dependence on the break-up time.

\vs
Let us now consider a quark density dependent 
hadronization cross section which is at least inversely proportional
to the density of quarks ($n \sim 1/V \sim t^{-3}$).
Exactly this assumption was made in transchemistry \cite{Biro:1999cx}.
Introducing the initial collison time,
\be
  t_{c} = \frac{1}{\langle \sigma v_0 \rangle \rho_0},
  \label{ORIG_COLL_TIME}
\ee
with density $\rho_0 = 2N_0/V_0$ we arrive at
\be
  \frac{\langle \sigma v_0 \rangle}{V_0} = \frac{1}{2N_0t_c}.
   \label{INIT_RATE}
\ee
The quantity $rN_0$ describing the deviation from equilibrium
for hadrons then simplifies to
\be
 rN_0 = \frac{t_0}{2t_c} \log \frac{t_{br}}{t_0}
 \label{DEV_RHIC}
\ee
for one-dimensional relativistic flow and
\be
 rN_0 = \frac{t_0}{6t_c} \left(
 1 - \left(\frac{t_0}{t_{br}}\right)^3 \right)
 \label{DEV_SPS}
\ee
for three-dimesional non-relativistic flow of massive constituents.
From these formulae one realizes that for $t_{br}=t_0$,
i.e. in case of immediate break-up the chemical equilibrium cannot
be approached, no hadron $C$ will be formed and $\gamma_C$ remains
zero. The opposite limit, $t_{br} \rightarrow \infty$ is less
uniform: while in the relativistic one-dimensional scenario 
equilibrium is always
reached if enough time is given ($\gamma_C(\infty)=1$), this is not
so for a stronger, more spherical expansion or in the presence of a
transverse expansion. 
In the case of three-dimensional spherical expansion
$\gamma_C$ tends to a finite value,
\be
 \gamma_C(\infty) = \frac{1}{1+6t_c/t_0},
\label{FINITE-GAMMA}
\ee
which is in general less than one. It means that 
{\em chemical equilibrium cannot be really reached}
even in infinite time. (An analysis of 
proton -- deuteron mixture with a similar conclusion was given in
Ref.\cite{Biro:1984yz}.) For early break-up, $t_{{\rm br}}-t_0 \ll t_0$,
both scenarios lead to
\be \gamma_C \approx  \frac{t_{br}-t_0}{2t_c} \ll 1 . \ee


\vs
In quark matter hadronization scenarios eq.(\ref{FINITE-GAMMA})
contradicts to the quark confinement principle. In order to
reach $\gamma_C(\infty)=1$ a dynamical confinement mechanism,
causing the quark fusion cross section in medium to be
(at least) inversely proportional to the quark density, has to
be taken into account besides the equilibrium suppression of
quark number (referred in this paper as ``static confinement'').

\vs
Let us assume that $A$ and $B$ are colored objects,
and the in-medium quark fusion cross section scales as
\be
\sigma = \sigma_0 \left( \frac{V}{V_0} \frac{2N_0}{N_A+N_B} \right)^{1+\epsilon}
\label{SIGMA-SCALES}
\ee
with some positive $\epsilon$. The solution of the rate equation
in the $R_{{\rm eq}}=0$ case (both static and dynamic confinement) becomes
\be
N_C = N_0 \left( 1 - (1-\delta)^{1/\epsilon} \right),
\ee
with
\be
\delta = \frac{t_0}{6t_c} \left( \left( \frac{t}{t_0} \right)^{3\epsilon}
- 1 \right)
\ee
in the 3-dimensional flow (SPS) scenario.
For $\epsilon = -1$ (without dynamical confinement) one gets back
the result discussed so far. In the limiting case $\epsilon=0$
(minimal dynamical confinement) one obtains
\be
N_C = N_0 \left( 1 - \left( \frac{t_0}{t} \right)^{\frac{t_0}{2t_c}} \right).
\label{TRANS-CHEM}
\ee
For a 1-dimensional flow (RHIC) scenario we arrive at
\be
N_C = N_0 \left( 1 - e^{-(t-t_0)/2t_c} \right)
\ee
in the $\epsilon = 0 $ case.

\vs 
Fig.2 plots the dependence of the non-equilibrium ratio 
 $N_C/N_0$ on the scaled break-up
time $t_{br}/t_0$ for these scenarii.


\begin{figure}
 \centerline{\rotate[r]{\psfig{figure=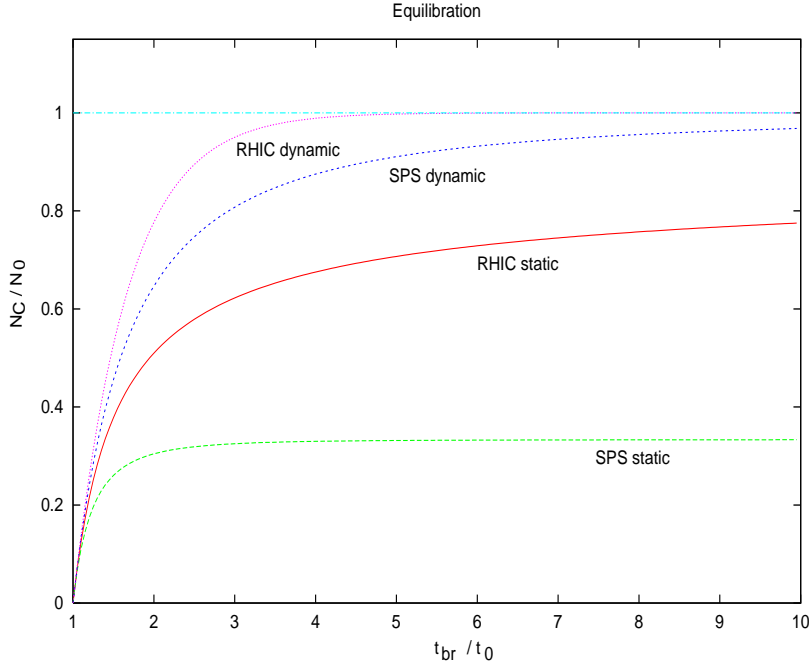,width=90mm,height=110mm}} }

\label{EQ-FIG}
\caption{
Degree of the chemical equilibrium reached for different
break-up times $t_{{\rm br}}/t_0$ at initial inter-collison times $t_c/t_0=1/3$
for SPS (3-dimensional)  and RHIC (1-dimensional) expansion scenarios 
with static and dynamic confinement (hadronization cross section is inversely
proportional to the color density).
}

\end{figure}



\vs
\section{Conclusion}

\vs
In conclusion
1) the results of the statistical model can be interpreted not only by 
assuming hadronic equilibrium, but as well by quark -- hadron mixture
chemistry with quarks eventually suppressed due to confinement. 
This interpretation
avoids the collision time problem: the hadrons need not transmutate into
each other directly.
2) The ALCOR and constrained hadronic equilibrium leads to the 
same result in the simple scenario presented here 
due to the constraints on conserved quantities.
Quark coalescence, which takes conservation into account, closely mimics
hadronic statistics in chemical equilibrium.
This feature does not generalize for several competing reaction
channels, but the overall picture is still similar. 
3) $\gamma_C \approx 0.3$ is typical for spherical expansion 
(CERN SPS) without considering dynamical confinement,
even for an arbitrary long hadronization process.
The experimental evidence, reaching at least 70-80\% of equilibrium
ratios for almost all hadrons in a short time in the order of
several fm/c, excludes this possibility. It also underlines
the quark matter hadronization scenario with dynamical confinement
effect, as transchemistry utilizes it. 
4) A purely longitudinal expansion scenario, 
perhaps typical for RHIC and LHC energies, in principle
allows for reaching the equilibrium even with static confinement only, 
but the process is rather slow and an early break-up may
prevent some quarks from hadronizing. Dynamical confinement has to be
in work for near one-dimensional expansion scenarios with a
minor transverse flow, too.
5) We pointed out that total elimination of free quarks in an expanding
quark matter scenario in the presence of any transverse flow
{\em requires} a quark density dependence of the hadronization
cross section, which is at least $1/n$. We call this effect ``dynamical''
confinement, it evidently cannot be of perturbative origin.

\vs
One is tempted to speculate about the dynamical confinement,
looking for a cause of the density dependence of the hadronization
cross section. Since this dependence makes the cross section stronger
and even diverging with diminishing quark density, 
it cannot have perturbative origin.
In our view it has to do with the formation of strings or color ropes,
which become longer and hence more efficiently bind quarks into
hadronic (color neutral) clusters as the 
color charge density drops.

\vs
{\bf Acknowledgements}
This work was supported by the US-Hungarian Joint Fund T\'eT 649 
and by the Hungarian National Research Fund OTKA T029158.


\vs



\end{document}